\documentclass[12pt]{article}
\usepackage{amsfonts,epsfig,latexsym,amscd}
\usepackage{amssymb}%,letterspace}

\newcommand{\R}{{\mathbb{R}}}

\newcommand{\I}{{\mathbb{I}}}

\newcommand{\be}{\begin{equation}}
\newcommand{\ee}{\end{equation}}
\newcommand{\bea}{\begin{eqnarray}}
\newcommand{\eea}{\end{eqnarray}}
\newcommand{\bean}{\begin{eqnarray*}}

\newcommand{\eean}{\end{eqnarray*}}
\font\upright=cmu10 scaled\magstep1

\newcommand{\PP}{\hbox{\upright\rlap{I}\kern 1.5pt P}}

\newcommand{\identity}{{\upright\rlap{1}\kern 2.0pt 1}}

\newcommand{\HH}{\mbox{\hbox{\upright\rlap{I}\kern 1.7pt H}}}

\newcommand{\fr}{\frac}

\newcommand{\ra}{\rightarrow}

\newcommand{\sg}{\sigma}

\newcommand{\pr}{\partial}
\newcommand{\x}{ {\bf x} }
\newcommand{\hs}{\hspace{5mm}}

\newcommand{\dg}{\dagger}

\newcommand{\ve}{\varepsilon}
\newcommand{\acc}{\\[3mm]}

\newcommand{\vv}{{\bf v}}

\begin{document}
\setcounter{page}{0}
\begin{titlepage}
\strut\hfill
\vspace{0mm}
\begin{center}

{\large\bf The Non-Compact Weyl Equation}
\vspace{12mm}

{\bf Anastasia Doikou${}^*$ \ and \ Theodora Ioannidou${}^\dg$}
\\[8mm]
\noindent ${}^*${\footnotesize Department of Engineering Sciences, University of Patras,
GR-26500 Patras, Greece }\\
{\footnotesize {\tt E-mail: adoikou@upatras.gr}}
\\[8mm]
\noindent ${}^\dg${\footnotesize Department of Mathematics, Physics and Computational Sciences, Faculty of Engineering,\\
Aristotle University of Thessaloniki, GR-54124 Thessaloniki, Greece }\\
{\footnotesize  {\tt E-mail: ti3@auth.gr}}

\vspace{12mm}

\begin{abstract}
\noindent
A {\it non-compact} version of the Weyl equation is proposed,
based on the infinite dimensional spin zero representation of the $\mathfrak{sl}_2$ algebra.
Solutions of the aforementioned equation are obtained in terms of the Kummer functions.
In this context,  we discuss the ADHMN approach in order to construct the corresponding {\it non-compact} BPS monopoles.
\end{abstract}

\end{center}
\end{titlepage}
\tableofcontents
\section{Introduction}

The Nahm equations provide a system of non-linear ordinary differential equations
\be \fr{dT_i}{ds}=\fr{1}{2}\, \ve_{ijk}\,[T_j,\ T_k]
\label{Nahm}
\ee
for three $n\times n$ anti-hermitian matrices $T_i$ (the so-called Nahm data) of complex-valued functions of the variable $s$, where $n$ is the magnetic charge of the BPS monopole configuration.
The tensor $\ve_{ijk}$ is the totally antisymmetric tensor.

In the ADHMN approach, the construction of  $SU(n+1)$ monopole solutions of the Bogomolny equation with topological charge $n$ is translated to the following problem which is known as
the inverse Nahm transform \cite{Nahm}.
Given the Nahm data for a $n$-monopole the one-dimensional Weyl equation
\be
\left( \I_{2n}\fr{d}{ds}-\I_n\otimes x_j \sg_j +iT_j\otimes\sg_j\right)\vv({\bf x},s)=0
 \label{Weyl}
\ee
for the complex $2n$-vector $\vv(\x,s)$, must be solved.
$\I_n$ denotes the $n\times n$ identity matrix, ${\bf x}=(x_1,x_2,x_3)$ is the position in space at which the monopole fields are to be calculated.
In the minimal symmetry breaking case, the Nahm data $T_i$'s can be cast as
(see Reference \cite{msbook}, for a more detailed discussion)
\be
T_i =-{i\over 2}\,f_i\,\tau_i, \hs i=1,\ 2,\ 3 \label{data}
\ee
where $\tau_i$'s form  the $n$-dimensional representation of $SU(2)$ and satisfy:
\be
[\tau_i,\ \tau_j] = 2i \varepsilon_{ijk}\,\tau_k.
\ee

Let us choose an orthonormal basis for these solutions, satisfying
\be
\int \hat \vv^\dg \hat \vv \,ds=\I. \label{wnor}\ee
Given $\hat \vv(\x,s)$, the normalized vector computed from (\ref{Weyl}) and (\ref{wnor}), the Higgs field $\Phi$ and the  gauge potential $A_i$ are given by
\bea
\Phi&=&-i\int s\, \hat \vv^\dg \hat \vv \,ds,\label{Higgs}\\
A_i&=&\int \hat \vv^\dg \,\pr_i \hat \vv\, ds.
\eea

In \cite{DI, DI2}, we applied the ADHMN construction to obtain the $SU(n + 1)$ (for generic values of $n$) BPS monopoles with minimal symmetry breaking, by solving the Weyl equation.
In this paper, we present a non-compact approach  of the ADHMN transform by introducing an infinite dimensional spin zero representation of the $\mathfrak{sl}_2$ algebra for the Nahm data.
The aforementioned representation is expressed in terms of appropriate {\it differential operators};  hence, the Weyl equation is also written in terms of the aforementioned differential operators, and not in terms of $n \times n$ matrices as in its conventional form (see, for example, Ref. \cite{DI, DI2}).
In the Appendix, we present the equivalence between the two approaches, i.e. matrix versus differential operator description of the Weyl equation, which leads us to conjecture that the results of the present investigation should by construction satisfy the Bogomolny equation. This is mainly due to the structural similarity between the equations arising in the present case, and the ones emerging in the {\it finite dimensional case} described in the Appendix and in Ref. \cite{DI}. Nevertheless, this is an intriguing issue, which merits further investigation, in particular when azimuthal dependence is also implemented along the lines described in \cite{DI2}.

\section{The Weyl Equation}

In order to construct the non-compact BPS monopole solutions of the Weyl equation, let us consider the $\mathfrak{sl}_2$ algebra, and focus on the non-trivial spin zero representation.

Consider the general case: i.e. the spin $S \in \R$ representation of $\mathfrak{sl}_2$ of the form
\be
\tau_1 =\! -\!\left(\xi^2 -1\right){d\over d\xi} + S \left(\xi +\xi^{-1}\right), \ \ \ \tau_2 = - i\left [\left(1+\xi^2\right) {d\over d\xi} + S\left(\xi^{-1} -\xi\right) \right ], \ \ \  \tau_3= -2\xi\, {d\over d\xi}.\ \label{rep1}
\ee
Also take the inner product, in the basis of polynomials of
$\xi$ on the unit circle ($\xi = e^{i\theta}$), to be of the form:
\be
\langle f, g \rangle\equiv {1\over 2i\pi} \int {1\over \xi} \,f^* g\, d\xi \label{inner}
\ee
and immediately obtain  the formula
\be
\langle \xi^m, \xi^n \rangle = \delta_{nm}. \label{inner2}
\ee

Next consider the generic state
\be
\vv = \sum_{k=-\infty}^{\infty} \,h_k\, \xi^k \left(b_1 \sqrt{\eta} + \fr{b_2}{\sqrt{\eta}}\right),
\ee
where $h_k=h_k(r,s)$ and $b_i=b_i(r,s)$ for $i=1,2$.

Notice that using the representation (\ref{rep1}), for $S$ being an {\it integer or half integer};
together with the inner product (\ref{inner}) of an appropriate orthonormal basis
$\{\hat{\vv}_1,\dots,\hat{\vv}_{n+1}\}$ where  $n=2S+1$ being the dimension of the representation (see also Appendix for more details):
\be
\int_0^{n+1}\,\langle \hat{\vv}_i,\hat{\vv}_j \rangle  \,ds = \delta_{ij}
\ee
one may recover the Higgs field obtained in \cite{DI} from the formula
\be
\Phi_{ij}= -i \int_0^{n+1} (s-n) \,\langle \hat{\vv}_i,\hat{\vv}_j \rangle  \,ds.  \label{higgs2}
\ee

Next, we focus  on the the {\it spin zero representation of $\mathfrak{sl}_2$}, associated to
the M\"obius transformation and also relevant in high energy QCD (see for example, Ref.  \cite{lipatov, FK}).
Again we consider the spherically symmetric case (that is, $x_i = r \delta_{i3}$)
where  the  Nahm data are given by (\ref{data}) for $f_i = f= -{1\over s}$.

Substituting  the Nahm data (\ref{data}) where $\tau_i$'s are defined by (\ref{rep1}) for $S=0$  to the Weyl
equation (\ref{Weyl}) and expressing $\sigma_i$ in terms of the spin $\fr{1}{2}$ representation; that is equation (\ref{rep1})  for $S=\fr{1}{2}$:
\be
\sigma_1=\! -\!\left(\eta^2-1\right) {d \over d\eta} + \fr{ \left(\eta^{-1} +\eta\right)}{2}, \ \
\sigma_2 = \!-i\left [\left(1+\eta^2\right) {d \over d\eta} + \fr{\left(\eta^{-1}-\eta\right)}{2}\right], \ \ \sigma_3 =-2\eta\, {d\over d\eta} \label{ss}
\ee
one gets
 \bea
&&\!\!\!\!\!\!\!\!\!\!\!\left\{ {d \over ds}  +{f \left(\xi^2 -1\right)\over 2} {d\over d\xi}
\left[\left(\eta^2-1\right) {d\over d\eta} - {\left(\eta^{-1} +\eta\right) \over 2}\right] -{f \left(1+\xi^2\right)\over 2}  {d\over d\xi}\left[\left(1+\eta^2\right) {d\over d\eta} + {\left(\eta^{-1} -\eta\right)\over 2}  \right] \right . \nonumber\acc
&&\!\!\!\!\!\!\!\!\!\!\!\left. + \,2f \xi \,{d \over d\xi}\left( \eta {d\over d\eta}\right) +2r \eta \,{d\over d\eta} \right\}
\sum_{k =-\infty}^{\infty} h_k \,\xi^k \left(b_1 \sqrt{\eta} + \fr{b_2}{\sqrt{\eta}} \right)
= 0.\label{g}
\eea

Next, by setting $w_k = b_1 \,h_k$ and $u_k = b_2 \,h_k$ in (\ref{g}), the following set of linear differential equations is obtained
\bea
&& \dot{w}_k -{\left(k+1\right) \over s}\, u_{k+1} - \left({k\over s} -r\right)w_k  =0, \nonumber\acc
&& \dot{u}_{k+1}+ {k\over s} \,w_k +\left(\fr{\left(k+1\right)}{s}-r\right) u_{k+1}=0, \hs \hs k \in (-\infty,\ \infty).
\eea
Here, $\dot{w}_k$ and $\dot{u}_k$ are the total derivatives of the functions $w_k$ and $u_k$ with respect to the argument $s$. Note that our results are analogous to the ones obtained in \cite{DI}.

Let us now solve these equations. The coupled equations for $u_{k+1}$ and $w_k$ are equivalent by expressing $u_{k+1}$ in terms of $w_k$:
\be
u_{k+1}={s\over \left(k+1\right)}\, \dot{w}_k - {\left(k-rs\right)\over \left(k+1\right)} \, w_k,\label{uk1}
\ee
 to the single second-order equation
\be
s \ddot{w}_k + 2 \dot{w}_k - \left[r^2s-2r \left(k+1\right)\right]w_k=0.\label{w}
\ee
Then, the solution of (\ref{w}) is given in a closed form, in  terms of the Kummer functions as
\be
w_k = e^{-rs}\Big[c_1(r)\, M\left(-k,2, 2rs\right) +c_2(r)\, U\left(-k, 2, 2rs\right)\Big]\label{sol}
\ee
where $c_i(r)$ for $i=1,2$ are constants.  $M\left(-k,2, 2rs\right) $ is the regular {\it confluent hypergeometric  Kummer function} and $U\left(-k,2, 2rs\right)$ is the {\it Tricomi confluent hypergeometric function} defined in Table $1$\footnote{$\Gamma(a,z)$ is the {\it  complementary} or {\it upper incomplete Gamma function} defined by
$$\Gamma(a,x)=\int_x^\infty t^{a-1}\,e^{-t}\,dt,\hs \Re(a)>0.$$}.
These functions are widely known as the Kummer functions of first and second kind, respectively, and are linearly independent solutions of the Kummer equation \cite{AS}.
\begin{center}
{\small \begin{tabular}{|r||r||l|}
\hline
 & $M\left(-k,2, 2rs\right) \hs\hs\hs\hs\hs$  &\,\,  $U\left(-k,2, 2rs\right) $  \acc
\hline \hline
$k=-2$  & $e^{2rs}$\,\,\, \hs\hs\hs\hs\hs\hs\hs\hs\hs&
$\Gamma\left(-1,2rs\right)e^{2rs}$\acc
$k=-3$ & $\left(1+rs\right)e^{2rs}$\hs\hs\hs\hs\hs\hs\hs&
 $\fr{1+2rs}{4rs}-\left(1+rs\right)\Gamma\left(0,2rs\right)e^{2rs}$\acc
$k=-4$& $\fr{1}{3}\left(3+6rs+2r^2s^2\right)e^{2rs}$\,\,\,\hs\hs\hs&
 $\fr{1+5rs+2r^2s^2}{12rs}-\fr{1}{6}\left(3+6rs+2r^2s^2\right)\Gamma\left(0,2rs\right)e^{2rs}$\acc
 $k=-5$& $\fr{1}{3}\left(3+9rs+6r^2s^2+r^3s^3\right)e^{2rs}$&
 $\fr{\left(3+2rs\right)\left(1+8rs+2r^2s^2\right)}{144rs}\!-\!\fr{1}{18}\left(3\!+\!9rs\!+\!6r^2s^2\!+\!r^3s^3\right)\Gamma\left(0,2rs\right)e^{2rs}$\acc
\hline
\end{tabular}\acc
{\bf Table 1:} Explicit expressions of the Kummer functions $M\left(-k,2, 2rs\right) $ and $U\left(-k,2, 2rs\right) $ for $k=-2,\dots,-5$.\acc}
\end{center}

Finally, the corresponding function $u_{k+1}$ given by (\ref{uk1}) takes the simple form
\be
u_{k+1}=\fr{k}{\left(k+1\right)}\,e^{-rs}\Big[-c_1(r)\,M(-k+1,2,2rs)\,+\,c_2(r)(k+1)\,U(-k+1,2,2rs)\Big].
\label{uk}
\ee

The next step is to choose an orthogonal basis of the infinite dimensional space.
Consider the following functions:
\be
\vv_{k} = \xi^k\sqrt{\eta} \, w_k+  \fr{\xi^k}{\sqrt{\eta}}\,u_{k+1} ,
\ee
which are orthogonal by construction. Then the norm of such a function is given by
\bea
\int_{-\infty}^{1}\!\!\!  <\vv_k, \vv_k>ds&=& \int_{-\infty}^{1}  \left( w_k^2 + u_{k+1}^2\right) ds\nonumber\acc
&=&{\cal N}_k.\label{Nk}
\eea
As it can be observed from  Table $1$ the arbitrary constant $c_2(r)$ at (\ref{sol}) and (\ref{uk}) should be set equal to zero in order to avoid the divergencies of   (\ref{Nk}) at $s\ra -\infty$. Also, the norm (\ref{Nk})  is well-defined only  for $k\in(-\infty,-2]$.

Some particular examples of the  values of the norm ${\cal N}_k$  are
{\small \bea
\!\!\!\!\!\!\!\!\!\! {\cal N}_{-2}\!\! &=&\!\! \fr{c_1^2(r)}{2r}\left(3+4r+4r^2\right)e^{2r},\nonumber\acc
 \!\!\!\!\!\!\!\!\!\!{\cal N}_{-3}\!\! &=&\!\! \fr{c_1^2(r)}{8r}\left(5+16r+28r^2+16r^3+4r^4\right)e^{2r},\nonumber\acc
\!\!\!\!\!\!\!\!\!\!{\cal N}_{-4}\!\! &=&\!\! \fr{c_1^2(r)}{162r}\left(63+324r+864r^2+960r^3+540r^4+144r^5+16r^6\right)e^{2r},\nonumber\acc
\!\!\!\!\!\!\!\!\!\! {\cal N}_{-5}\!\! &=&\!\! \fr{c_1^2(r)}{288r}\left(81+576r+2088r^2+3456r^3+3084r^4+1536r^5+432r^6+64r^7+4r^8\right)e^{2r}.
\eea}

Similarly to the finite case the associated Higgs field may be then obtained via the generic
expression:
\be
\Phi_{kk}=-\fr{i}{{\cal N}_k}\,\int_{-\infty}^{1} s \left(w_k^2 + u_{k+1}^2\right) ds.
\ee

\section{Conclusions}

In this paper, we discuss the ADHMN construction in the case of the non-compact  $\mathfrak{sl}_2$ algebra.
More precisely, we propose a generalized version of the Weyl equation  in terms of differential operators.
The aforementioned  (non-compact) Weyl equation is solved explicitly for the infinite dimensional spin zero representation of $\mathfrak{sl}_2$, and the associated solutions are expressed in terms of the
so-called Kummer functions.
Also, a suitable infinite set of orthogonal functions is chosen, and in analogy to the finite case
(see, for example, \cite{DI} and References therein), expressions of the relevant Higgs fields are proposed.
These expressions have a simple and elegant form, and should correspond to a kind of infinite BPS monopole configurations.

The next natural step is to verify that our results satisfy the Bogomolny equation. However in order to do so we need  to implement azimuthal dependence to the spherically symmetric solution presented here along the lines presented in \cite{DI2} and thus,  obtain the solution of the full  non-compact  Weyl equation. This requires the identification of a suitable transformation \cite{DI2} that reduces the full problem to the ``diagonal'' one treated here.
This is arguably a highly non-trivial task, and will be pursued in full detail elsewhere together with the physical description of the full solution.
In any case,  the results presented here are already of
great significance given that they provide solutions of the {\it non-compact}  Weyl
equation, opening also the path to the study of novel infinite type monopole
configurations.

To conclude, it would be interesting to investigate any possible relevance of our findings with previous results of the classical version of the Nahm equations related to infinite monopoles \cite{W1,GCP} and $SU(\infty)$ Yang-Mills theories \cite{FIT,FFZ}.
Note that in \cite{W2} the Nahm equations are associated to the {\it classical}  $\mathfrak{sl}_2$ algebra ({\it Poisson bracket} structure) and are {\it linear}, whereas in our study we consider the {\it quantum}  $\mathfrak{sl}_2$ algebra and we deal with the corresponding spin zero infinite dimensional (non-compact) representation.
One final comment is in order: one should not confuse the spin zero infinite dimensional representation utilized here with the $n \to \infty$ limit of the representation used, for example, in \cite{DI, DI2}. The representation employed in the present investigation is qualitatively different from the $n \to \infty$ case;  having, for instance, a zero Casimir (i.e. $C=0$) as opposed to the $n \to \infty$ case where $C \to \infty$.

\appendix
\section{Appendix}

In what follows we  briefly describe the equivalence between the matrix description of the Weyl equation presented in full generality in Ref. \cite{DI} and the differential operator description attempted here.
More precisely, we  express the Weyl equation in terms of differential operators via the spin $S$ representation of $\mathfrak{su}_2$, for $S$ being  {\it integer} or {\it half integer}.
That is, we focus on the finite representation of dimension $n =2S +1$ and explicitly show the equivalence with the generic results obtained in \cite{DI}, via the matrix description.

First, consider the Nahm data (\ref{data}) with $f_i = -{1\over s}$; substitute the representations $\tau_i$ via the generic expressions (\ref{rep1}) for $S$  integer or half integer, and the matrices $\sigma_i$ via (\ref{ss}). Then the Weyl equation takes the form:
\bea
&&\!\!\!\!\!\!\!\!\!\!\!\!\!\!\!\!\!\!\!\! \left\{ {d\over d s} - {1\over 2s} \left [(\xi^2-1) {d \over d \xi} -S \left(\xi +\xi^{-1}\right)\right ] \!\!\!\left [ \left(\eta^2 -1\right){d\over d \eta} -\fr{\left(\eta^{-1} + \eta\right)}{2}\right ] \right.+ \nonumber\\
&&\!\!\!\!\!\!\!\!\!\!\!\!\!\!\!\!\!\!\!\! \left. {1\over 2s} \left [\left(\xi^2 +1\right) {d \over d \xi} +S \left(\xi^{-1} - \xi\right)\right ]\!\!\!\left [\left(\eta^2 +1\right) \fr{d}{d\eta}+\fr{\left(\eta^{-1} -\eta\right)}{2} \right ] \!\! -{2  \over s} \xi {d \over d \xi}\ \Big ( \eta {d\over d \eta}\Big )+2r \eta {d \over d \eta} \right \} \vv=0. \, \label{ww}
\eea
Finally, assume  the generic form for the function $\vv$ :
\be
\vv =\sum_{k=1}^n h_k\, \xi^{k-1-S} \left(b_1 \sqrt{\eta}+ {b_2\over  \sqrt{\eta}}\right),
\ee
where $h_k=h_k(r,s)$ and $b_i=b_i(r,s)$ for $i=1,2$.
Setting $w_k=b_1h_k$ and $u_k=b_2h_k$ in   (\ref{ww}),   the following set of linear differential equations is obtained
\bea
&& \dot{u}_1 - \left ( {n-1 \over 2s} + r \right) u_1 =0,  \label{1} \\
&& \dot{u}_{k+1} +{k-n\over s}\,w_k - \left ( {n-1-2k \over 2s} + r\right )u_{k+1} =0, \label{2}\\
&& \dot{w}_k - { k\over s} \,u_{k+1} + \left ( {n+1 -2k \over 2s}  + r \right )w_k=0, \label{3} \\
&& \dot{w}_n + \left ( {1 -n \over 2s} +r \right ) w_n =0, \label{4}
\eea
where $\dot{u}_i$ and $\dot{w}_i$ are the total derivatives of $u_i(r,s)$ and $u_i(r,s)$ with respect to the argument $s$.
Equations (\ref{1}) and (\ref{4}) can be immediately integrated and their solutions are equal to:
\be
u_1 = \kappa_1(r)\, s^{{n-1 \over 2}}\,e^{rs},\hs \hs \hs w_n = \kappa_2(r) \,s^{{n-1 \over 2}} \,e^{-rs}.
\ee
Note that the aforementioned solutions  coincide with the ones found in \cite{DI}.

The coupled equations (\ref{2}) and  (\ref{3}) are equivalent by expressing $u_{k+1}$ in terms of $w_k$:
\be
u_{k+1} ={1\over k} \left [s \dot{w}_k + \left ( {n+1 -2k \over 2} + rs \right )w_k \right] ,
\ee
to the single second-order equation
\be
s^2\ddot{w}_k + 2 s \dot{w}_k - \left[ r^2 s^2 +\left(n-1-2k\right)rs +{n^2-1 \over 4}\right] w_k =0,
\ee
which may be solved by substituting $w_k ={W_k \over s}$ and $z=2rs$. The latter equation is then reduced to the familiar Whittaker equation:
\be
{d^2W_k \over dz^2} + \left (-{1\over 4} + {2k-n+1 \over 2 z} + {1-n^2  \over 4z^2}\right ) W_k=0,
\ee
and coincides with the solution $W_k$ found in \cite{DI}.

The next step is to choose an orthogonal basis of the $n$-dimensional space.
Consider the following functions
\bea
 \vv_1 &=& {\xi^{-S}\over \sqrt{\eta}}\ u_1,  \nonumber\\
 \vv_{k} &=& \xi^{k-1-S}\left(\sqrt{\eta} \, w_k +  \fr{1}{\sqrt{\eta}}\,u_{k+1}\right), \nonumber\\
 \vv_n&=& \xi^{n-1-S}\sqrt{\eta}\ w_n, \hs \hs\hs \hs\hs \hs\hs \hs k\in \{2, \ldots, n-1 \}
\eea
which are  orthogonal by construction.  Then the norm of such a function is given by
\bea
\int_{0}^{n+1}\!\!\!  <\vv_k, \vv_k>ds&=& \int_{0}^{n+1}  \left( w_k^2 + u_{k+1}^2\right)ds\nonumber\acc
&=&{\cal N}_k.\label{Nk2}
\eea
And one may readily recover the Higgs field obtained in  \cite{DI}  from the formula
\be
\Phi_{kk}= -{i\over {\cal N}_k}  \int_0^{n+1} \left(s-n\right) \left(w_k^2 + w_{k+1}^2\right)  \,ds.  \label{higgs22}
\ee
{\bf Remark}: It is clear that the present description is equivalent to the one discussed in \cite{DI}.


\begin{thebibliography}{99}
\vspace{.25cm}

\bibitem{Nahm}
\noindent W. Nahm, {\it The construction of all self-dual multimonopoles by the
ADHM method}, in {\it Monopoles in Quantum Field Theory}, eds N.S. Craigie, P. Goddard and
W. Nahm (World Scientific, Singapore, 1982).

\bibitem{msbook}
N.S. Manton and P.M. Sutcliffe, {\it Topological Solitons},
Cambridge Monographs on Mathematical Physics, Cambridge University Press (2004).

\bibitem{DI}
A. Doikou and T. Ioannidou, JHEP 1008, (2010) 105.

\bibitem{DI2}
A. Doikou and T. Ioannidou, {\tt arXiv:1010.5076}.

\bibitem{lipatov}
L. N. Lipatov, Sov. Phys. JETP 63, 904 (1986).

\bibitem{FK}
L.D. Faddeev and G.P. Korchemsky, Phys. Lett. B342, 311 (1995).

\bibitem{AS}
M. Abramowitz and I. Stegun, {\it Handbook of Mathematical Functions with Formulas,
Graphs and Mathematical Tables}, New York Dover (1972).

\bibitem{W1}
R.S. Ward,Phys. Lett. B 234, 81 (1990).

\bibitem{W11}
R.S. Ward,  Class. Quantum Grav. 7, L95 (1990);
Class. Quantum Grav. 7, L217 (1990);

\bibitem{GCP}
H. Garcia-Compean and J.F. Plebanski, Phys. Lett. A234, 5 (1997).

\bibitem{FIT}
E.G. Floratos, J. Iliopoulos and G. Tiktopoulos, Phys. Lett. B217, 285 (1989).

\bibitem{FFZ}
D.B. Fairlie,  P. Fletcher and C.K. Zachos, J. Math. Phys. 31, 1088 (1990).

\bibitem{W2}
R.S. Ward,  J. Geom. Phys. 8, 317 (1992).

\end{thebibliography}
\end{document}